\documentclass{epl}

\usepackage{amssymb}
\usepackage{amsmath}
\usepackage{amsfonts}

\title{Optimal strategies in collective Parrondo games}
\shorttitle{Optimal strategies in Parrondo games}

\author{Luis Dinis  \and Juan M.R.~Parrondo}

\institute{Grupo Intedisciplinar de Sistemas Complejos (GISC) and
Dept.~de F\'{\i}sica At\'{o}mica,
 Molecular y Nuclear,
 Universidad
Complutense de Madrid, 28040-Madrid, Spain.}

\begin{document}

 \maketitle
\begin{abstract}
We present a modification of the so-called Parrondo's paradox
where one is allowed to choose in each turn  the game that a large
number of individuals play. It turns out that, by choosing the
game which gives the highest average earnings at each step, one
ends up with systematic loses, whereas a periodic or random
sequence of choices yields a steadily increase of the capital. An
explanation of this behavior is given by noting that the
short-range maximization of the returns is ``killing the goose
that laid the golden eggs".  A continuous model displaying similar
features is analyzed  using dynamic programming techniques from
control theory.
\end{abstract}

The physics of Brownian motors has recently inspired the discovery
of a counterintuitive phenomenon in gambling games, which is
attracting considerable attention. Ajdari and Prost showed that a
one dimensional Brownian particle in a flashing asymmetric
potential experiments a net motion in a given direction
\cite{ajdari}. The motion persists even against a small force.
Consequently, one can have the following startling situation: the
particle moves in the direction of the force if the potential is
on or if it is off, whereas it moves in the opposite direction if
the potential is flashing. The translation of the dynamics of this
Brownian particle to gambling games constitutes the so-called {\em
Parrondo's paradox} \cite{nature,fnl}: two losing games yield,
when alternated, a winning game. The effect is obtained with two
games, A and B, that mimic the behavior of the Brownian particle
in a flat and a ratchet potential, respectively.

In game A, the capital $X(t)$ of the player increases by one unit
with probability $p_1=1/2-\epsilon$, where $\epsilon$ is a small
positive real number, and decreases by one unit with probability
$1-p_1$. In the following, we will interpret the game as a bet on
the toss of a slightly biased coin.

Game B is played with two coins depending on $X(t)$: if $X(t)$ is
not a multiple of three, we use coin 2, with a probability to win
 $p_2=3/4-\epsilon$ and a probability to lose $1-p_2$; if $X(t)$
is a multiple of three, we use coin 3, with a probability to win
$p_3=1/10-\epsilon$
 and a probability to lose $1-p_3$. It can be proved that the combination of the
  ``good" coin 2 and the ``bad" coin 3 yields a fair game when
$\epsilon=0$, and a losing game when $\epsilon>0$ \cite{fnl}. By
fair, losing and winning here we mean that the average capital
$\langle X(t)\rangle$ is a constant, decreasing or increasing
function of $t$, respectively.

We then have two games, A and B, which are fair (losing) if
$\epsilon=0$ ($\epsilon>0$). The aforementioned counterintuitive
effect is that the alternation of A and B, in some given random or
periodic sequences, is a winning game.

The phenomenon indicates that the alternation of stochastic
dynamics can result in a behavior which differs qualitatively from
that exhibited by each of the dynamics, and therefore, could in
principle be relevant in a variety of situations, ranging from
economics to physics, where the constrains or the dynamics of a
system switches between two arrangements \cite{fnl}.

However, the paradox loses all its interest if one is allowed to
choose the game to play in each turn. In this case, the trivial
strategy is to choose A when $X(t)$ is a multiple of three and B
otherwise. The resulting game is clearly winning and this strategy
performs better than any periodic or random alternation of games A
and B. We could call these latter strategies ``blind", since they
do not use any information about the state of the system.

In a similar way, if one has some information about the position
of the particle in a flashing ratchet, it is possible to switch on
and off the potential in such a way that energy is extracted from
a single thermal bath. This is nothing but a Maxwell demon
\cite{review}.

For a Brownian particle, it is known that the acquisition of this
information, or its subsequent erasure from a memory device, has
some unavoidable entropy cost \cite{leff}, which prevents any
violation of the Second Law of Thermodynamics. On the other hand,
in other contexts,  like economics, there is no such limitations
and it is unlikely that blind strategies could be of any interest.

However, the model that we present in this Letter shows that this
is not the case. It is a modification of the original Parrondo's
paradox in which blind strategies are winning whereas a strategy
which chooses the game with the highest average return is losing.
Moreover, we will identify the mechanism underlying this
counterintuitive behavior and show with a second model that it can
also appear in simple deterministic systems.

The two models presented  are also of interest in control theory.
The choice of a strategy maximizing some quantity is a problem
widely treated by optimal control theory, which has been proven a
powerful tool in a number of disciplines including engineering,
physics, chemistry, economics, social sciences, medicine and
biology \cite{lay,lew}. The counterintuitive phenomenon discussed
in this Letter, up to our knowledge, has not been reported before
and can be relevant in optimization problems involving dynamical
systems.

The first model consists of a large number $N$ of players. At each
turn of the game, a fraction $\gamma$ of these players is randomly
selected. We are told how much money every player has and we are
then allowed to choose a game, A or B, which will be played by
\emph{all} $\gamma N$ players  in the subset. Our goal is to
choose each turn between A or B in order to maximize the average
earnings of the players. We consider three different strategies:
\begin{itemize}
\item \emph{Periodic strategy}: the game is selected by
 following a given periodic sequence, for example
ABBABB$\ldots$
\item \emph{Random strategy}:  the game is chosen
randomly with equal probability for both A and B.
\item
\emph{Short-range (SR) optimal strategy}: the game that
 will yield the highest average return is chosen.
\end{itemize}

\begin{figure}
\begin{center}
\includegraphics[width=6cm]{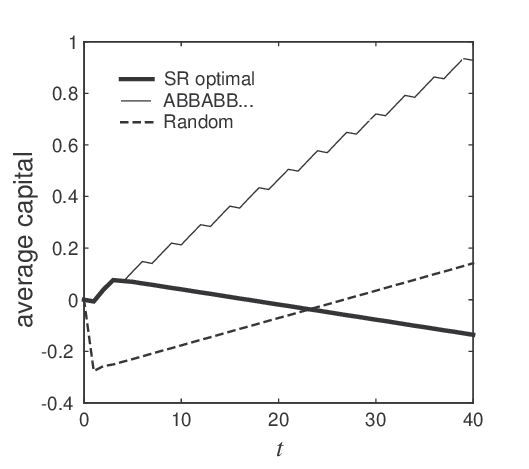}
\caption{Evolution of the average money of an infinity number of
players for $\gamma=0.675$ and $\epsilon=0.005$ and the three
strategies discussed in the text.} \label{figura2}
\end{center}
\end{figure}

As we will see below, the third strategy makes use of the
available information whereas the periodic and random strategies
are blind, in the sense defined above. Surprisingly, these blind
strategies produce a systematic winning whereas the SR optimal
strategy is losing, as it is shown in figure \ref{figura2}.

A detailed analysis of our model will reveal the underlying
mechanism causing this unexpected phenomenon. The key magnitude
for this analysis is $\pi_0(t)$, the fraction of players whose
money is a multiple of three in turn $t$. From $\pi_0(t)$, it is
not difficult to calculate the average fraction of players that
would win in each game:
\begin{eqnarray}  p_{\mbox{winA}} &=& p_1
 \nonumber\\
 p_{\mbox{winB}}&=&\pi_0p_3+(1-\pi_0)p_2
\end{eqnarray}
The SR optimal strategy chooses the game which gives the highest
return in one turn. Comparing $p_{\mbox{winA}}$ and
$p_{\mbox{winB}}$ we get the following prescription:
\begin{eqnarray}
\mbox{play A} &\mbox{if}&  \pi_{0}(t)\geq
\pi_{0c}\nonumber\\
\mbox{play B} &\mbox{if} & \pi_0(t) < \pi_{0c}
\label{prescription}
\end{eqnarray}
with
 $\pi_{0c}\equiv(p_1-p_2)/(p_3-p_2)=5/13$.

Let us focus now on the behavior of $\pi_0(t)$ for $\epsilon=0$.
On one hand, $\pi_0(t)$ tends to 1/3 if A is played a large number
of turns, because, under the rules of A, the capital $X(t)$ is a
symmetric and homogenous random walk. On the other hand, if B is
played repeatedly, $\pi_0(t)$ tends to 5/13, i.e., to $\pi_{0c}$.
This can be proved by analyzing game B as a Markov chain
\cite{fnl}. Notice also that this coincidence was expected since B
is fair game for $\epsilon=0$ and $\pi_{0c}$ has been obtained by
solving $p_{\rm winB}=p_{\rm winA}=1/2$.

\begin{figure}[h]
\begin{center}
\includegraphics[width=7cm]{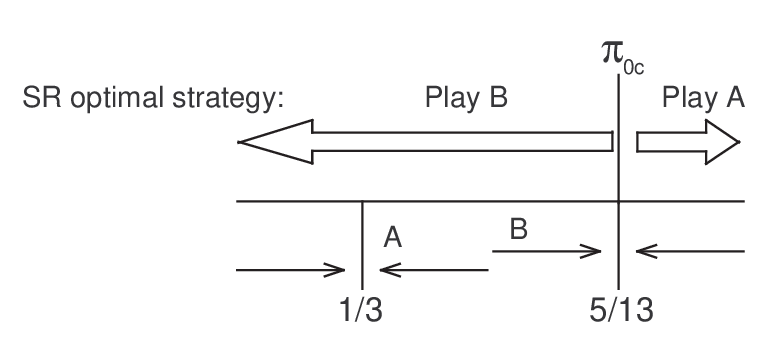}
\caption{Schematic representation of the evolution of $\pi_0(t)$
under the action of game A and game B. The prescription of the SR
optimal strategy is
 also
represented.}\label{scheme}
\end{center}
\end{figure}

Figure  \ref{scheme} represents schematically the evolution of
$\pi_0(t)$ under the action of each game, as well as the
prescription of the SR optimal strategy given by
Eq.~\eqref{prescription}.  Now we are ready to explain why the SR
optimal strategy yields worse results than the periodic and random
sequences.

\begin{figure}
%\twoimages[width=6cm]{fig3.eps}{fig4.eps}
\twoimages[height=6cm]{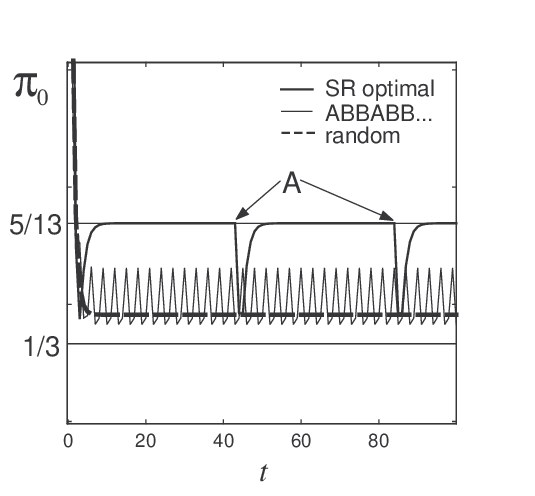}{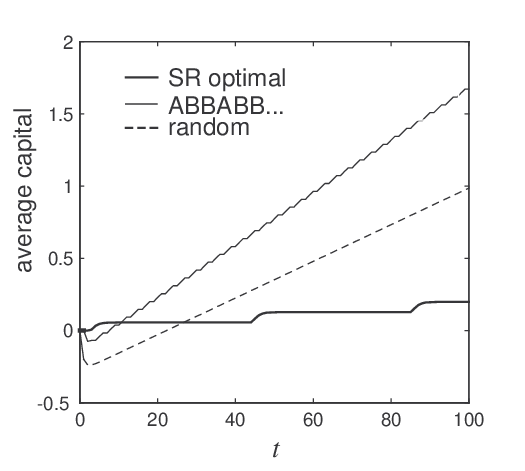}
 \caption{Evolution of $\pi_0$ (left) and the average capital (right) for
$N=\infty$, $\gamma=0.5$, $\epsilon=0$ and the three different
strategies. The arrows in the left figure show the turns where the
short-range optimal strategy chooses game A and they coincide with
the steps of the curve representing the
capital.}\label{figgamma05}
\end{figure}

Consider an initial distribution of the capital such that
$\pi_0(0)<\pi_{0c}=5/13$. The SR optimal strategy chooses B and,
consequently, $\pi_0$ increases. If $\pi_0(1)$ is still under 5/13
(and this is the case for $\gamma$ small enough), the SR optimal
strategy chooses B again. We see that, as far as $\pi_0(t)$ does
not exceed 5/13, the SR optimal strategy chooses B in every turn.
However, this choice, although it is the one which gives the
highest returns in each turn, drives $\pi_0(t)$ towards 5/13,
i.e., towards values of $\pi_0(t)$ where the gain is small. For
instance, if $\gamma=1/2$, the SR optimal strategy chooses B forty
times in a row before switching to game A. This will make $\pi_0$
approximately equal to $\pi_{0c}=5/13$ at almost every turn, as
can be seen in  figure \ref{figgamma05} (left). The same figure
(right)  shows that, as long as $\pi_0(t)$ is close to $\pi_{0c}$,
the average capital remains approximately constant.

On the other hand, the random and the periodic strategies choose
game A even when $\pi_0<\pi_{0c}$. This will not produce earnings
in this specific turn, but will take $\pi_0$ away from $\pi_{0c}$
and make the corresponding average money grow faster than that for
the SR optimal strategy, as can be  seen in the figure.

 In other words, the SR optimal strategy, by choosing B
too many times, is ``killing the goose that laid the golden eggs",
and to perform better than this strategy one must sacrifice
short-term profits for higher returns in the future, as the blind
strategies considered here do.

The introduction of $\epsilon$ has two consequences: it turns A
and B into losing games when played alone, and it decreases the
stationary value for game B, which now is {\em below}
$\pi_{0c}=5/13$ (the critical value $\pi_{0c}$ does not change).
Due to the latter, the system may get trapped playing game B
forever when following the SR optimal strategy and, since B is now
a losing game, the average money will decrease. This is, for
instance, the situation already presented in figure \ref{figura2}
for $\epsilon=0.005$ and $\gamma=0.675$.

Now we present a continuous and deterministic model which displays
some of the features of the previous one. Consider the following
dynamical system:
\begin{eqnarray}
 \dot y(t)&=&\alpha (t) x(t) \nonumber\\
 \dot   x(t)&=&-\frac{1}{\tau}\left[ x(t)- x_{\rm fc}(1-\alpha(t))\right]
 \label{eq:sistema}
\end{eqnarray}
with $\alpha(t)=0$  or  1. Our task is to find $\alpha(t)$ that
maximizes $y(T)$. These equations are a rather generic model of a
system which produces some output like, for instance, a production
plant. $y(t)$ is the total cumulative output of the plant up to
time $t$. We can decide to switch on and off the plant at every
time $t$ by setting $\alpha(t)=1$ or $\alpha(t)=0$, respectively.
Finally, $x(t)$ is the productivity of the plant, which decreases
exponentially when the plant is working and goes back to its full
capacity value $x_{\rm fc}$ when the plant is off, $\tau$ being
the characteristic time of these relaxations. If we are allowed to
use the plant up to a time $T$, the problem is to find the
protocol or {\em policy} $\alpha(t)$ maximizing the total output
$y(T)$.

 A naive approach to the problem consists of maximizing $\dot y(t)$
at every time $t$:
\begin{eqnarray}
\alpha(t)=
    \begin{cases}
      0& \mbox{if }x(t)<0 \\
      1& \mbox{if }x(t)\geq 0
    \end{cases}
     \label{eq:condicion2}
\end{eqnarray}
However, it is not hard to see that this set up will keep
$y(t)=y(0)$ for all $t$
 if initially the productivity $x(0)$ is negative,
or make $y(t)$ tend to $y(0)+\tau x(0)$ exponentially if $x(0)>0$.
In either case, $y(t)$ will attain a saturation value. The
criterion \eqref{eq:condicion2} prescribes making the plant work
whenever the productivity is positive, and this is equivalent to
the short-range optimal strategy in our previous model, which
dictated to play game B if $\pi_0(t)<\pi_{0c}$. If the
productivity is positive, we of course get more when the plant is
working, but then we also get a decrease of the productivity which
will end up exhausting the system. With the prescription given by
\eqref{eq:condicion2} we are also killing the goose that laid the
golden eggs, and it is again possible to do better by letting the
plant rest even when the productivity $x(t)$ is positive.

Surprisingly enough and despite the linearity and simplicity of
our system, the precise optimal policy $\alpha(t)$ is not easy to
find. Some of the techniques provided by control theory fail in
this case, such as the Euler-Lagrange equations and the Pontryagin
principle \cite{lay}. The only way to completely solve the problem
is to optimize a discrete version such as: \begin{eqnarray}
x_{k+1} &= & x_k-h(x_k-1+\alpha_k)\nonumber
\\ y_{k+1} &=&
y_k+h\alpha_kx_k 
\end{eqnarray}
where $h$ is a small time step, and we have taken $x_{\rm
fc}=\tau=1$ for simplicity. The optimization of this discrete
system can be done applying the so-called {\em Bellman's
Optimality Criterion} \cite{lay}, which states that in the optimal
policy, the final decisions $\alpha_k$ ($k=n,\dots,N$)  are
optimal {\em given} the state resulting from the first decisions,
$\alpha_k$ ($k=0,\dots,n-1$). That is, we can find the
$\alpha_{N-n}$ maximizing $J_n\equiv y_{N+1}-y_{N-n}$ as a
function of $x_{N-n}$, $y_{N-n}$, recursively from $n=0$ to $n=N$
\cite{dinis}.

\begin{figure}[t]
\begin{center}
\includegraphics[width=6cm]{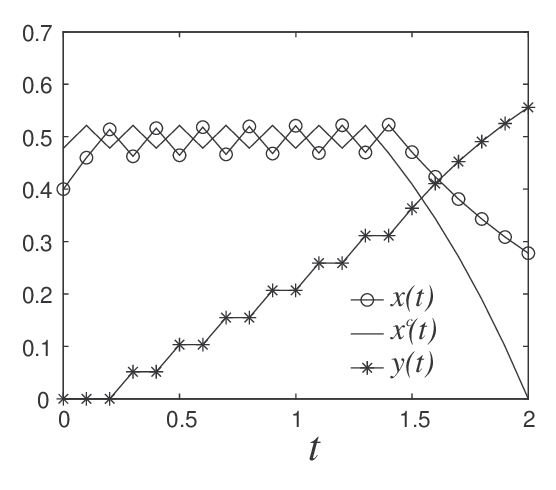}
\caption{The cumulative output $y(t)$, the productivity $x(t)$,
and the critical value $x^c(t)$ for the discrete version of the
system \eqref{eq:sistema}, $T=2$, $h=0.1$, and $\tau=x_{\rm fc}=1$.} \label{figura9}
\end{center}
\end{figure}

\begin{figure}
\begin{center}
\includegraphics[width=6cm]{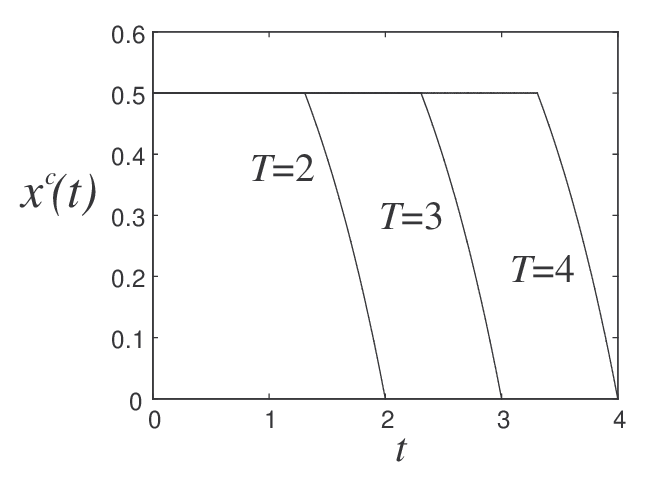}
\caption{Numerical computation of $x^c(t)$, for $h=0.001$, $T=2$,
3 and 4, and $\tau=x_{\rm fc}=1$.} \label{figura10}
\end{center}
\end{figure}

The optimal choice of $\alpha(t)$ in terms of the state $x(t)$
happens to be:
\begin{equation}
\alpha(t)=
    \begin{cases}
      0& \mbox{if }x(t)<x^c(t) \\
      1& \mbox{if }x(t)\geq x^c(t)
    \end{cases}
    \label{eq:leycontrol}
\end{equation}
where $x^c(t)$ is a critical value for the productivity which can
be calculated by solving the resulting recurrence equations. In
figure \ref{figura9} we show $x^c(t)$ for $\tau=x_{\rm fc}=1$,
$h=0.1$, and $T=2$, as well as the behaviour of $y(t)$ and $x(t)$
(we have chosen a relatively big time step to make clear the rapid
changes in $\alpha(t)$ and $x(t)$ ). We see that this optimal
policy achieves a steadily increase of the output $y(t)$. Fig.
\ref{figura10} shows a numerical computation of $x^c(t)$ now for a
shorter time step $h=0.001$, and different values of the total
time, $T=2$, 3, and 4. Here we can see again that $x^c(t)=0.5$
until the final part of the interval $[0,T]$.

Consequently, the behaviour of the optimal policy $\alpha(t)$ is
as follows. There is a first stage in which the productivity
$x(t)$ goes to 0.5 by setting $\alpha=0$ if $x(0)<0.5$ and
$\alpha=1$ if $x(0)\ge 0.5$. Once $x(t)$ reaches the value 0.5,
the optimal policy prescribes very rapid changes between
$\alpha=0$ and $\alpha=1$ which keep $x(t)\simeq 0.5$. Finally, in
the last part of the interval $[0,T]$, when $x^c(t)$ starts to
decrease, the optimal policy chooses $\alpha(t)=1$.

The first two stages can be easily explained, even for arbitrary
values of $\tau$ and $x_{\rm fc}$. For this purpose, let us assume
for a moment that $\alpha(t)$ is constant but can take any value
within the interval $[0,1]$. Then $x(t)$ reaches a stationary
value $x_{\rm st}=x_{\rm fc}(1-\alpha)$. This gives the following
stationary slope for the cumulative output $y(t)$:
\begin{equation}
\dot y(t)=\alpha x_{\rm st}=x_{\rm
fc}\left[\alpha(1-\alpha)\right]
\end{equation}
which is maximum for $\alpha=0.5$ and  $x_{\rm st}=x_{\rm fc}/2$.
Therefore, the optimal policy should try to drive $x(t)$ to
$x_{\rm fc}/2$ and keep it there, i.e., should try to make the
plant work at half of its full capacity. In our case, $\alpha(t)$
can only take values 0 and 1. However, by rapid oscillations one
can get an effective value of $\alpha(t)$ equal to any real number
between 0 and 1. The optimal policy implies a rapid variation
which gives an effective value $\alpha=0.5$, yielding a slope for
$y(t)$ which is $\dot y(t)=x_{\rm fc}/4$, definitely better than
the short-range optimization of $\dot y(t)$ which gave us a
horizontal slope.

The final stage of the optimal policy $\alpha(t)$ has a clear
intuitive explanation. We have to abandon the plant at time $T$.
Therefore, the optimal policy when $t$ is approaching $T$ should
set $\alpha(t)=1$, since we want to get as much as possible and do
not care if we leave the plant exhausted after $T$. This is
exactly what happens to a middle-distance runner: she keeps a
constant velocity which allows her to maintain a stationary regime
but she sprints in the last meters of the race to use up all her
strength. With this picture in mind, we call this last stage the
``sprint". One can calculate the duration of the sprint, $t_{\rm
sprint}$, in our model, assuming that $x(0)=x_{\rm fc}/2$,
$\alpha(t)=1/2$ for $t<T-t_{\rm sprint}$ and $\alpha(t)=1$ for
$t>T-t_{\rm sprint}$. Eq.~\eqref{eq:sistema} can then be fully
solved yielding: \begin{equation} y(T) = \frac{x_{\rm
fc}}{4}\left( T-t_{\rm sprint}\right) +  \frac{\tau x_{\rm
fc}}{2}\left( 1 - e^{-t_{\rm sprint}/\tau}\right)
\end{equation}
>From this expression one easily finds that $y(T)$ reaches its
maximum for $t_{\rm sprint}=\tau\ln 2\simeq 0.693\,\tau$, in
agreement with the curves in Fig.~\ref{figura10}.

In conclusion, we have presented a stochastic model in which a
short-range optimization yields to systematic loses, whereas blind
strategies steadily win. We have found an explanation of this
phenomenon based on the fact that the short-range optimal strategy
is ``killing the goose that laid the golden egg", and proven that
the same mechanism can also arise in a linear deterministic
system. In fact, similar mechanisms has been widely reported in
the realm of economics and ecology, although mainly described
qualitatively. The risks of overtaxing commerce and the overuse of
natural resources are representative cases. We believe that the
models presented here could inspire new quantitative approaches to
these problems.

\acknowledgements

We are grateful to C. Van den Broeck for fruitful discussions.
This work has been financially supported by grant
BFM2001-0291-C02-02 from Ministerio de Ciencia y Tecnolog\'{\i}a
(Spain) and by a grant from {\em  Del Amo Program} (Universidad
Complutense).

%--------------------REFERENCIAS-----------------------------

\end{document}